\documentclass[12pt]{article}

\newcommand{\be}{\begin{equation}}
\newcommand{\ee}{\end{equation}}
\newcommand{\bi}[1]{\vspace{-3mm} \bibitem{#1}}
\usepackage{epsfig,amsmath,amssymb,graphics,graphicx}

\begin{document}

%%%\today

\begin{center}
{\Large \bf Fractional Generalization of Kac Integral}
\vskip 5 mm

{\large \bf Vasily E. Tarasov$^{1,2}$, and George M. Zaslavsky$^{1,3}$ }\\

\vskip 3mm

{\it $1)$ Courant Institute of Mathematical Sciences, New York University \\
251 Mercer St., New York, NY 10012, USA }\\ 
{\it $2)$ Skobeltsyn Institute of Nuclear Physics, \\
Moscow State University, Moscow 119992, Russia } \\
%%{E-mail: tarasov@theory.sinp.msu.ru}}
{\it $3)$ Department of Physics, New York University, \\
2-4 Washington Place, New York, NY 10003, USA } 
\end{center}

\begin{abstract}
Generalization of the Kac integral and Kac method for paths measure based on 
the L\'evy distribution has been used to derive fractional diffusion equation. 
Application to nonlinear fractional Ginzburg-Landau equation is discussed.
\end{abstract}

%%%%%%%%%%%%%%%%%%%%%%%%%%%%%%%%%%%%%%%%%%%%%%%%%%%%%%%%%%%%%%%%%%%%%%%%%%%%%%

\section{Introduction}

Kac integral \cite{Kac0,Kac,Moral} appears as a path-wise presentation of Brownian motion and 
shortly becomes, with Feynman approach \cite{F}, 
a powerful tool to study different processes described by the wave-type  
or diffusion-type equations.
In the basic papers \cite{Kac0,F}, the paths distribution 
was based on averaging over the Wiener measure. 
It is worthwhile to mention the Kac comment that the Wiener measure can be replaced by 
the L\'evy distribution that has infinite second and higher moments. 
There exists a fairly rich literature related to functional integrals 
with generalization of the Wiener measure (see for example \cite{Levy1,Levy2}). 
Recently the L\'evy measure was applied to derive a fractional generalization of 
the Schr\"odinger equation \cite{Laskin,Laskin2} using the Feynman-type approach and 
expressing the L\'evy measure through the Fox function \cite{Fox}

In this paper, we derive the fractional generalization of the diffusion equation (FDE) 
from the path integral over the L\'evy measure using the integral equation approach of Kac.

\section{L\'evy distribution}

Let us consider the transition probability $P(x,t|x',t')$ that 
describes the evolution of the probability density $\rho(x,t)$ by the equation
\be
\rho(x,t)=\int^{+\infty}_{-\infty} dx' \, P(x,t|x',t') \rho(x',t') ,
\ee
where
\be
\int^{+\infty}_{-\infty} dx \rho(x,t)=1 .
\ee
The function  $P(x,t|x',t')$ can be considered as conditional distribution function.
Then the normalization condition
\be
\int^{+\infty}_{-\infty} dx P(x,t|x',t')=1 
\ee 
holds.
Assume that $P(x,t|x',t')$ satisfies the Markovian (semigroup) condition
\be
P(x,t|x_0,t_0)=\int^{+\infty}_{-\infty}dx' P(x,t|x',t')P(x',t'|x_0,t_0) 
\ee
known also as the Chapman-Kolmogorov equation.

In physical theories, the stability of a family of probability 
distributions is an important property which basically states that 
if one has a number of random variables that belong to some family, 
any linear combination of these variables will also be in this family. 
The importance of a stable family of probability distributions is that they 
serve as "attractors" for linear combinations of non-stable random variables. 
The most noted examples are the normal Gaussian distributions, 
which form one family of stable distributions. 
By the classical central limit theorem the linear sum of a set of 
random variables, each with a finite variance, tends to the 
normal distribution as the number of variables increases.
All continuous stable distributions can be specified by the proper 
choice of parameters in the L\'evy skew alpha-stable distribution \cite{Levy} 
that is defined by
\be
L(x,y,\alpha,\beta,c)=\frac{1}{2\pi} \int^{+\infty}_{-\infty} 
dp \, e^{-ipx} U(p,y,\alpha,\beta,c) ,
\ee
where
\be \label{LD}
U(p,y,\alpha,\beta,c)
=\exp \Bigl(  iyp- |cp|^{\alpha} [1-i \beta sign(p) \Phi(\alpha,p) ] \Bigr)  ,
\ee
and
\be 
\Phi(\alpha,p)=
\begin{cases} 
\tan (\pi \alpha /2) , & 0< \alpha \le 2 , \quad \alpha \ne1 ;
\cr 
-(2/\pi)\log|p| , & \alpha=1 .
\end{cases}
\ee
Here $y$ is a shift parameter, $\beta$ is a measure of asymmetry, with $\beta=0$ 
yielding a distribution symmetric about $y$. 
In Eq. (\ref{LD}), parameter $c$ is a scale factor, which is a measure of the width of 
the distribution and $\alpha$ is the exponent or index of the distribution.

Consider $P(y,t'|x,t)$ as a symmetric homogeneous L\'evy  alpha-stable distribution
\be \label{Pyx}
P(y,t'|x,t) \equiv K(y-x,t'-t)=
\frac{1}{2\pi}  
\int^{+\infty}_{-\infty} dp \exp \Bigl( ip(y-x)- (t'-t) C_{\alpha} |p|^{\alpha}  \Bigr) ,
\quad (0<\alpha \le 2) .
\ee
For $\alpha=2$, Eq. (\ref{Pyx}) gives the Gauss distribution
\be
P(y,t'|x,t)= \frac{1}{\sqrt{4 \pi C_2 (t'-t)}} 
\exp \Bigl( - \frac{1}{4C_2(t'-t)} (y-x)^2 \Bigr) .
\ee

Eq. (\ref{Pyx}) gives the function
\be \label{K}
K(x,t)= \frac{1}{2\pi}  
\int^{+\infty}_{-\infty} dp \exp \Bigl( ip x- t C_{\alpha} |p|^{\alpha}  \Bigr) 
\ee
that can be presented as a Fourier transform
\be \label{KF}
K(x,t)= {\cal F}^{-1} \Bigl( e^{- t C_{\alpha} |p|^{\alpha}} \Bigr) ,
\ee
where
\be
{\cal F}^{-1}( f(p) ) = \frac{1}{2\pi}  
\int^{+\infty}_{-\infty} dp \, e^{ip x} f(p) .
\ee
For $\alpha=2$, Eq. (\ref{KF}) gives
\be \label{Ka2}
K(x,t)= \frac{1}{\sqrt{4 \pi C_2 t}} 
\exp \Bigl( -\frac{x^2}{4C_2 t} \Bigr) .
\ee
In the general case, the function $K (x,t)$, given by Eq. (\ref{KF}), 
can be expressed in terms of the Fox $H$-function 
\cite{Laskin,Laskin2,Fox,Mathai,Srivastava,West2,GN} (see Appendix).

%%%%%%%%%%%%%%%%%%%%%%%%%%%%%%%%%%%%%%%%%%%%%%%%%%%%%%%%%%%%%%%%%%%%%%%%%%
\newpage
\section{Fractional Kac path integral}

Let us denote by $C[t_a,t_b]$ the set of trajectories starting 
at the point $x_a=x(t_a)$ at the time $t_a$ and having the endpoint
$x_b=x(t_b)$ at the time $t_b$.

%%%%%%%%%%%%%%%%%%%%%%%%%%%%%%%%%%%%%%%

The Kac functional integral \cite{Kac,Moral,CD} is 
\be \label{FKI}
W(x_b,t_b|x_a,t_a)= \int_{C[t_a,t_b]}  {\cal D}_W x(t) \,  
\exp \Bigl( - \int ^{t_b}_{t_a}d \tau V(x(\tau)) \Bigr) ,
\ee
where $V(x)$ is some function, and
\be \label{Dwx}
{\cal D}_W x =\lim_{n \rightarrow \infty} 
\prod^{n}_{k=1} K(\Delta x_k,\Delta t_k) d x_k .
\ee
For (\ref{Ka2}), expression (\ref{Dwx}) gives  
\be \label{DwxN}
{\cal D}_W x =
\lim_{n \rightarrow \infty} \prod^{n}_{k=1} 
\frac{dx_k}{\sqrt{4\pi C_2 \Delta t_k}} 
\exp \Bigl( -\frac{(\Delta x_k)^2}{4C_2 \Delta t_k} \Bigr) ,
\ee
which is the Wiener measure of functional integration \cite{CD}.
The integral (\ref{FKI}) is also called the Feynman-Kac integral.
Using (\ref{K}) for $\alpha=2$, the path integral (\ref{FKI}) can be written as
%%%The functional integral (\ref{36}) is defined as the limit
\be \label{psFK}
W(x_b,t_b|x_a,t_a)=\lim_{n \rightarrow \infty}
\frac{1}{(2\pi)^n} \int_{\mathbb{R}^{2n}} dx_1 \, dp_1 \, ... \, dx_n \, dp_n 
\exp \sum^n_{k=0} 
\Bigl( ip_k \Delta x_k - \Delta t_k [ C_2 p^2_k +V(x_k)] \Bigr) ,
\ee
where the time interval $[t_a,t_b]$ is partitioned as
\be
t_k=t_a+k\frac{t_b-t_a}{n}, \quad t_0=t_a, \quad t_n=t_b ,
\ee
and
\be
\Delta x_k =x_{k+1}-x_k, \quad \Delta t_k =t_{k+1}-t_k , 
\quad x_k=x(t_k), \quad p_k=p(t_k) .
\ee
The functional integral (\ref{psFK}) can be rewritten as
\be \label{psFK2}
W(x_b,t_b|x_a,t_a)=\int {\cal D} x {\cal D} p \, \exp \Bigl( 
\int^{t_b}_{t_a} dt \Bigl[ ip \dot{x} -  C_{\alpha} p^2 - V(x) \Bigr] \Bigr).
\ee
where
\be \label{DxDp}
{\cal D} x =\lim_{n \rightarrow \infty} \prod^{n}_{k=1} d x_k , \quad
{\cal D} p =\lim_{n \rightarrow \infty} \prod^{n}_{k=1} \frac{dp_k}{2\pi} .
\ee
The Kac functional integral in the form (\ref{psFK2}) is 
a classical analog of the Feynman phase-space path integral, 
which is also called the path integral in Hamiltonian form. \\

For the fractional generalization of Wiener measure (\ref{Dwx}) 
and Kac integral (\ref{FKI}), we consider $K(x,t)$ given by (\ref{K}).
Substitution of (\ref{K}) into
\be  \label{Apr1}
W(x_b,t_b|x_a,t_a)=\lim_{n \rightarrow \infty} 
\int_{\mathbb{R}^n} \prod^{n}_{k=1} dx_k \,  
K(\Delta x_k,\Delta t_k) \exp \Bigl( - \Delta t_k V(x_k) \Bigr) 
\ee
with
\be
K(\Delta x_k,\Delta t_k)=\frac{1}{2\pi} \int^{+\infty}_{-\infty} dp_k \exp 
\Bigl( ip_k \Delta x_k- \Delta t_k  C_{\alpha} |p_k|^{\alpha}  \Bigr) ,
\quad (0<\alpha \le 2) ,
\ee
gives
\be \label{25}
W(x_b,t_b|x_a,t_a)=\lim_{n \rightarrow \infty} \int_{\mathbb{R}^{2n}} 
\prod^{n}_{k=1} \frac{dx_k \, dp_k}{2 \pi} \, \exp \sum^n_{k=0} 
\Bigl( ip_k \Delta x_k - \Delta t_k [ C_{\alpha}|p_k|^{\alpha}  +V(x_k)] \Bigr) .
\ee
Similarly to (\ref{psFK2}), (\ref{DxDp}) this expression can be written as
\be \label{FFKI}
W(x_b,t_b|x_a,t_a)=\int {\cal D} x {\cal D} p \, \exp \Bigl( 
\int^{t_b}_{t_a} dt \Bigl[ ip \dot{x} -  C_{\alpha}|p|^{\alpha}  - V(x) \Bigr] \Bigr).
\ee
This expression is a fractional generalization of (\ref{psFK2}). 

If we introduce formally imaginary time such that
\[ i \dot{x}=i\frac{dx}{dt}=\frac{dx}{ds}, \]
then (\ref{FFKI}) transforms into the Feynman path integral 
with a generalized action \cite{Laskin,Laskin2}
\[ S[x,p]=   \int^{t_b}_{t_a} dt \Bigl[  p \dot{x} -  C_{\alpha}|p|^{\alpha}  - V(x) \Bigr] \]
as an action.
Hamiltonian-type formal equations of motion are
\be
\frac{dx}{ds}=N_{\alpha} |p|^{\alpha-1}, 
\quad
\frac{dp}{ds}=-\frac{\partial V(x)}{\partial x} ,
\ee
where $N_{\alpha}=\alpha C_{\alpha} \, sign(p)$.

%%%%%%%%%%%%%%%%%%%%%%%%%%%%%%%%%%%%%%%%%%%%%%%%%%

\section{Fractional diffusion equations}

It is known that the Kac integral (\ref{FKI}) can be considered 
as a solution of the diffusion equation \cite{Kac,CD}. 
Let us derive the corresponding diffusion equation for 
the fractional generalization of the Kac integral (\ref{FFKI}).

In (\ref{FFKI}) the integration is performed  over a set 
$C[t_a,t_b]$ of trajectories that start at point $x_a=x(t_a)$ 
at time $t_a$ and end at point $x_b=x(t_b)$ at time $t_b$.
For simplification, $t_a=0$, $x_a=0$, and $t_b=t$, $x_b=x$ are used. 
In particular, we can consider two following cases of $C[t_a,t_b]$. 

(1) The set $C_f [0,t]$ consists of paths for which 
both the initial and final points are fixed.
The integration over this set obviously gives 
the transition probability
\[ \int_{C_f [t_a,t_b]} {\cal D}_W x= K(x_b-x_a,t_b-t_a)=P(x_b,t_b|x_a,t_a) , \]
or
\[ \int_{C_f [0,t]} {\cal D}_W x= K(x,t) . \]
The conditional fractional Wiener measure corresponds to the integration over 
the set $C_f [0,t]$ of paths with fixed endpoints: $x_a=0$, $x_b=x$.
%%%The conditional measure is directly related to the transition probability.

(2) If we consider a set $C_a[0,t]$ of trajectories with arbitrary endpoint $x_b=x$,
the measure is called the unconditional fractional Wiener measure.
This measure satisfies the normalization condition
\be \label{29}
\int_{C_a[0,t]} {\cal D}_W x=
\int^{+\infty}_{-\infty} dx \, 
\int_{C_f [0,t]}  {\cal D}_W x =
\int^{+\infty}_{-\infty} dx \, K(x,t)= 1,
\ee
since it is a probability that the system ends up anywhere.

For simplification, we introduce the notation
\be \label{Z0}
Z[x,t] =\exp \Bigl( - \int^{t}_{0}d \tau V(x(\tau)) \Bigr) , 
\ee
and define the field
\be \label{Z}
u(x,t)=W(x,t|0,0) .
\ee 
For the fractional Kac functional integral, we have with respect to (\ref{29}), 
\be \label{FKI2}
\int_{C_a [0,t]}  {\cal D}_W x \,  Z[x,t]=
\int^{+\infty}_{-\infty} dx \, 
\int_{C_f [0,t]}  {\cal D}_W x \,  Z[x,t] . 
\ee

Using notations (\ref{Z0}), (\ref{Z}), expression
(\ref{FFKI}) for $t_a=0$, $x_a=0$, and $t_b=t$, $x_b=x$ can be presented as
\be \label{u} 
u(x,t)= \int_{C_f[0,t]} {\cal D}_W x \,  
\exp \Bigl( - \int ^{t}_{0}d \tau V(x(\tau)) \Bigr) = 
\int_{C[0,t]} {\cal D}_W x Z[x,t] .
\ee
To derive a fractional diffusion equation, we use the identity \cite{CD}
\be \label{id}
\exp \Bigl( - \int ^{t}_{0}d \tau V(x(\tau)) \Bigr) =
1- \int^t_0 d\tau \Bigl[ V(x(\tau)) \exp \Bigl( - \int ^{\tau}_{0} d s V(x(s) \Bigr) \Bigr] .
\ee
Equation (\ref{id}) can be proved by using differentiation by $t$, and 
the value of the constant is found from the condition of coincidence of 
both sides for $t=0$.
For the notation (\ref{Z}), identity (\ref{id}) has the form
\be \label{46}
Z[x,t]=1-\int^t_0 d\tau \Bigl[ V(x(\tau)) Z[x,\tau] \Bigr] .
\ee
Equation (\ref{46}) can be integrated with respect to the 
conditional fractional Wiener measure:
\be \label{Zeq}
\int_{C_f[0,t]} {\cal D}_W x \, Z[x,t]=
\int_{C_f[0,t]} {\cal D}_W x \, 1 \,  -
\int_{C_f[0,t]} {\cal D}_W x \, \int^t_0 d\tau \Bigl[ V(x(\tau)) Z[x,\tau] \Bigr] .
\ee
Changing the order of the integration in the second term in 
the right hand-side of (\ref{Zeq}), we get
\[
\int_{C_f[0,t]} {\cal D}_W x \, \int^t_0 d\tau \, \Bigl[ V(x(\tau)) Z[x,\tau] \Bigr] =
\int^t_0 d\tau \, \int_{C_f[0,t]} {\cal D}_W x \, \Bigl[ V(x(\tau)) Z[x,\tau] \Bigr] =
\]
\[
=\int^t_0 d\tau \, \int^{+\infty}_{-\infty} dx_{\tau} \, \int_{C_f[0,\tau]} {\cal D}_W x \,
\int_{C_f[\tau,t]} {\cal D}_W x \, \Bigl[ V(x(\tau)) Z[x,\tau] \Bigr] =
\]
\be \label{Second}
= \int^t_0 d\tau \, \int^{+\infty}_{-\infty} dx_{\tau} \, V(x(\tau)) \,
\int_{C_f[0,\tau]} {\cal D}_W x \, Z[x,\tau] \int_{C_f[\tau,t]} {\cal D}_W x .
\ee
The first term in the right hand-side of (\ref{Zeq}) gives
\[
\int_{C_f[t_a,t_b]} {\cal D}_W x \, 1 =
\lim_{n \rightarrow \infty} \int_{\mathbb{R}^n} \prod^{n}_{k=1} dx_k \,  
K(\Delta x_k,\Delta t_k) = \]
\be \label{First}
=\lim_{n \rightarrow \infty} \int_{\mathbb{R}^n} \prod^{n}_{k=1} dx_k \,  
K(\Delta x_k,\Delta t_k) = K(x_b-x_a,t_b-t_a) .
\ee

Using (\ref{Z}), (\ref{First}), and (\ref{Second}), Eq. (\ref{Zeq}) gives
the integral equation
\be \label{uu}
u(x,t)=K(x,t) - \int^t_0 d\tau \, \int^{+\infty}_{-\infty} dx_{\tau} \, V(x_{\tau}) \,
u(x_{\tau},\tau) K(x-x_{\tau},t-\tau) .
\ee
For this equation there exists the infinitesimal operator ${\cal L}_{\alpha}$ 
(generator) of time shift such that
\be
\frac{\partial u(x,t)}{\partial t}={\cal L}_{\alpha} u(x,t) .
\ee
%%%This generator can be defined by
%%%\be {\cal L}_{\alpha}=\lim_{t\rightarrow 0} \frac{}{} \ee

Using (\ref{u}) and (\ref{KF}), we obtain
\be \label{54}
{\cal L}_{\alpha}u(x,t)= C_{\alpha} \frac{\partial^{\alpha}}{\partial |x|^{\alpha}} u(x,t)
 -\lim_{t \rightarrow 0} \frac{1}{t}  \int^t_0 d\tau \int^{+\infty}_{-\infty} dy  
K(x-y,t-\tau) V(y) u(y,\tau) ,
\ee
where ${\partial^{\alpha}}/{\partial |x|^{\alpha}}$ 
is a fractional Riesz derivative \cite{SKM,OS,Podlubny,KST} of order $0<\alpha<2$ 
that is defined by its Fourier transform
\be
\frac{\partial^{\alpha}}{\partial |x|^{\alpha}} u(x,t)= 
{\cal F}^{-1}\Bigl( |p|^{\alpha} \tilde u(p,t)\Bigr)=
\frac{1}{2\pi} \int^{+\infty}_{-\infty} dp \, |p|^{\alpha} \tilde u(p,t) \, e^{-ipx} ,
\ee
where
\be
\tilde u(p,t)= \int^{+\infty}_{-\infty} dx \, u(x,t) \, e^{ipx} .
\ee
The initial condition $K(x,0)=\delta(x)$ gives \cite{CD}
\be \label{42}
\lim_{t \rightarrow 0}
\frac{1}{t}  \int^t_0 d\tau \int^{+\infty}_{-\infty} dy  
K(x-y,t-\tau) V(y) u(y,\tau) =V(x)u(x,t) .
\ee
Then (\ref{54}) gives
\be 
{\cal L}_{\alpha}= C_{\alpha} \frac{\partial^{\alpha}}{\partial |x|^{\alpha}} -V(x) .
\ee
This generator is an operator of fractional differentiation of order $\alpha$.

As a result, we obtain 
\be \label{FLE}
\frac{\partial u(x,t)}{\partial t}= 
C_{\alpha} \frac{\partial^{\alpha} u(x,t)}{\partial |x|^{\alpha}} - V(x)u(x,t) ,
\ee
which is a diffusion equation with fractional coordinate derivatives. 
For $\alpha=2$, Eq. (\ref{FLE}) is the usual diffusion equation.

It is worthwhile to mention that the way of obtaining fractional equation (\ref{FLE}) 
is based on the exploiting the properties of integral equation (\ref{uu}),
while the expansion of exponents in (\ref{25}) 
over small $\Delta t_k$ has been used in \cite{Laskin,Laskin2}
for Feynman path integral.

%%%%%%%%%%%%%%%%%%%%%%%%%%%%%%%%%%%%%%%%%%%%%%%%%%%%%%%%%%%%%%%

\section{Fractional diffusion equations by Kac approach}

It is useful also to derive the fractional diffusion equation 
from (\ref{FKI}) using Kac approach described in Sec. 4. of \cite{Kac}.

The mathematical expectation value of $Z[x,t]$ is defined as
\be \label{W1}
E \left< \exp \Bigl( - \int ^{t}_{0}d \tau V(x(\tau)) \Bigr) \right>=
\int_{C_a [0,t]} {\cal D}_w x \, \exp \Bigl( \int ^{t}_{0}d \tau V(x(\tau)  \Bigr) .
\ee
Using the expansion
\be
\exp \Bigl( - \int ^{t}_{0}d \tau V(x(\tau)) \Bigr) =
\sum^{\infty}_{m=0} \frac{(-1)^m}{m!} \Bigl( \int ^{t}_{0} d \tau V(x(\tau)) \Bigr)^m ,
\ee
we get 
\be \label{W1b}
E \left< \exp \Bigl( - \int ^{t}_{0}d \tau V(x(\tau)) \Bigr) \right>=
\sum^{\infty}_{m=0} \frac{(-1)^m}{m!}
\int_{C_a [0,t]} {\cal D}_W x \, \Bigl( \int ^{t}_{0} d \tau V(x(\tau)  \Bigr)^m .
\ee
The expression (\ref{W1b}) can be presented as
%%%(see Sec. IV of \cite{Kac}) as
\be
E \left< \exp \Bigl( - \int ^{t}_{0}d \tau V(x(\tau)) \Bigr) \right>=
\sum^{\infty}_{m=0} (-1)^m \int^{+\infty}_{-\infty} dx \, Q_m(x,t) , 
\ee
where 
\be \label{QQm}
Q_m(x,t) = \frac{1}{m!} \int_{C_f [0,t]} {\cal D}_W x \,
\Bigl( \int ^{t}_{0} d \tau V(x(\tau)) \Bigr)^m  .
\ee
These functions (\ref{QQm}) satisfy the recurrence equations \cite{Kac}
\be \label{Qm}
Q_{m+1}(x,t)=\int^{t}_0 d \tau \int^{+\infty}_{-\infty} dy \, K(x-y,t-\tau) V(y) Q_m(y,\tau) ,
\ee
and
\be \label{Q0}
Q_0(x,t)=K(x,t) .
\ee
Let us introduce
\be \label{51}
Q(x,t)= \sum^{\infty}_{m=0} (-1)^m Q_m(x,t) .
\ee
Then
\[ 
Q(x,t) = \sum^{\infty}_{m=1} \frac{(-1)^m}{m!} \int_{C_f [0,t]} {\cal D}_W x \,
\exp \Bigl( \int ^{t}_{0} d \tau V(x(\tau)) \Bigr)^m =
\]
\be
=\int_{C_f [0,t]} {\cal D}_W x \, \exp \Bigl( \int ^{t}_{0}d \tau V(x(\tau)  \Bigr) ,
%%%=u(x,t) ,
\ee
and
\be
E \left< \exp \Bigl( - \int ^{t}_{0}d \tau V(x(\tau)) \Bigr) \right>=
 \int^{+\infty}_{-\infty} dx \, Q(x,t) .
\ee
It follows from (\ref{Qm}) and (\ref{Q0}) that
the field $Q(x,t)$ satisfies the integral equation
\be
Q(x,t)=Q_0(x,t)- \int^t_0 d\tau \int^{+\infty}_{-\infty} dy  
K(x-y,t-\tau) V(y) Q(y,\tau) . 
\ee

%%%Semigroup property allows us to $Q(x,t)$ 

There exists an infinitesimal operator  ${\cal L}_{\alpha}$ 
of time shift such that
\be
\frac{\partial Q(x,t)}{\partial t}={\cal L}_{\alpha} Q(x,t) .
\ee
Using (\ref{Qm}) (\ref{QQm}), and (\ref{KF}),
this generator can be expressed through a fractional differential operator
\be
{\cal L}_{\alpha}Q(x,t)= C_{\alpha} \frac{\partial^{\alpha}}{\partial |x|^{\alpha}} Q(x,t)
 -\lim_{t \rightarrow 0} \frac{1}{t}  \int^t_0 d\tau \int^{+\infty}_{-\infty} dy  
K(x-y,t-\tau) V(y) Q(y,\tau) .
\ee
The initial condition $K(x,0)=\delta(x)$ gives similar to (\ref{42}) 
\be
\lim_{t \rightarrow 0}
\frac{1}{t}  \int^t_0 d\tau \int^{+\infty}_{-\infty} dy  
K(x-y,t-\tau) V(y) Q(y,\tau) =V(x)Q(x,t) .
\ee
%%%Then 
%%%\be {\cal L}_{\alpha}= C_{\alpha} \frac{\partial^{\alpha}}{\partial |x|^{\alpha}} -V(x), \ee
%%%where ${\partial^{\alpha}}/{\partial |x|^{\alpha}}$ 
%%%is a fractional Riesz derivative of order $0<\alpha<2$ \cite{SKM,OS,Podlubny,KST} 
%%%that is defined as Fourier transform of $|p|^{\alpha}$.

As a result, we obtain 
\be \label{QQQ}
\frac{\partial Q(x,t)}{\partial t}= 
C_{\alpha} \frac{\partial^{\alpha} Q(x,t)}{\partial |x|^{\alpha}} - V(x)Q(x,t) ,
\ee
which is fractional diffusion equation that coincides with (\ref{FLE}).
Then 
\be
Q(x,t)=W(x,t|0,0)=\int_{C_f[0,t]} {\cal D}_W x \, 
\exp \Bigl( - \int^{t}_{0} d\tau V(x(\tau)) \Bigr).
\ee
Using (\ref{51}), the approximate solution of (\ref{FLE}) can be presented as
\[
u(x,t) \approx Q_0(x,t)-Q_1(x,t)+Q_2(x,t)= \] 
\[
=K(x,t)-  \int^{t}_0 d \tau \int^{+\infty}_{-\infty} dy \, K(x-y,t-\tau) V(y) K(y,\tau) +
\]
\be
+\int^{t}_0 d \tau \int^{\tau}_0 dt' \int^{+\infty}_{-\infty} dy \, \int^{+\infty}_{-\infty} dy' \, 
K(x-y,t-\tau) V(y) \, K(y-y',\tau-t') V(y') K(y',t') .
\ee
for small enough $V(x)$.

\section{Nonlinear fractional equations}

Equations (\ref{FLE}) and (\ref{QQQ}) are linear equations with respect to 
the fields $u(x,t)$ and $Q(x,t)$.
In general, nonlinear equations can be derived 
from the functional integral over the space of branching paths 
(see \cite{Dal} and Sec. VI.4. of \cite{DF}). 
Note that Feynman path integral over the branching paths 
has been suggested in \cite{MC} (see also \cite{JPS,Peres}).
The multiplicative representations of nonlinear diffusion equations 
are also considered in \cite{Ch,Mar,Sev}. 
As an example of nonlinear diffusion equation, which can be derived from
integrals over the branching paths, is an equation with
the polynomial nonlinearity \cite{DF,Dal}:
\be
U(u)=\sum^m_{k=2} a_k [u(x,t)]^k .
\ee
Using fractional Kac integral over the branching L\'evy paths \cite{GJ,VH}, 
a nonlinear generalization of fractional equation (\ref{FLE}) 
can be derived in the form
\be 
\frac{\partial u(x,t)}{\partial t}= 
C_{\alpha} \frac{\partial^{\alpha} u(x,t)}{\partial |x|^{\alpha}} - V(x)u(x,t)
+\sum^m_{k=2} a_k [u(x,t)]^k .
\ee
For example, fractional equations with cubical nonlinearity can be obtained
\be \label{FGLE}
\frac{\partial u(x,t)}{\partial t}= 
C_{\alpha} \frac{\partial^{\alpha} u(x,t)}{\partial |x|^{\alpha}} - 
V(x)u(x,t) + a_3 [u(x,t)]^3 .
\ee
Equation (\ref{FGLE}) is the fractional 
generalization of the Gross-Pitaevskii equation \cite{Gross,Pit}.
For $V(x)=const$, Eq. (\ref{FGLE}) is fractional Ginzburg-Landau equation
that is suggested in \cite{WZ} (see also \cite{TZ1,Mil}) 
to describe complex media with fractional dispersion law.

\section*{Acknowledgments}

This work was supported by the Office of Naval Research, 
Grant No. N00014-02-1-0056, and the NSF Grant No. DMS-0417800.

%%%%%%%%%%%%%%%%%%%%%%%%%%%%%%%%%%%%%%%%%%%%%%%%%%%%%%%%%%%%%%%%%%%%%%%%%%%%%%%%%

%%%%%%%%%%%%%%%%%%%%%%%%%%%%%%%%%%%%%%%%%%%%%%%%%%%%%%%%%%%%%%%%%%%%%%%%%%%%%%%
\newpage
\section*{Appendix: Fox function representation for $K(x,t)$}

In this section, we use the results of the paper \cite{Laskin} (see also \cite{Laskin2})
to demonstrate how the function $K (x,t)$ defined by Eq. (\ref{K}) 
can be expressed in the terms of the Fox $H$-function 
\cite{Fox,Mathai,Srivastava,West2,GN}. 
The Fox function representation of $K(x,t)$ can be considered as 
a fractional analog of expression (\ref{Ka2}).  
%%%Note that derivatives and integrals of Fox function can be calculated 
%%%by formally manipulating with the parameters in the $H$-function. 
To present $K(x,t)$ in terms of the Fox $H$-function, 
we consider the Mellin transform of (\ref{K}). 
Comparing of the inverse Mellin transform with the
definition of the Fox function , 
we obtain an expression in terms of Fox $H$-function.

Using the relation $K(x,t )=K(-x,t )$,
it is sufficient to consider $K(x,t )$ for $x\geq 0$ only.
The Mellin transformation of (\ref{K}) is
\be
\stackrel{\wedge }{K}(s,t )=\int^\infty_0 dx \, x^{s-1} K(x,t )=
\frac{1}{2\pi} \int^\infty_0 dx \, x^{s-1}
\int^{+\infty}_{-\infty} dp \,
 \exp \Bigl( i px - C_\alpha |p|^\alpha t \Bigr) . 
\ee
Changing the variables
\[
p\rightarrow \left( C_{\alpha} t \right)^{-1/\alpha } \eta  ,
\qquad x \rightarrow \left( C_\alpha t \right)^{1/\alpha }\xi , 
\]
we present $\stackrel{\wedge}{K}(s,t )$ as
\be \label{eq20}
\stackrel{\wedge}{K}(s,t )=\frac{1}{2\pi} 
\left( (C_\alpha t )^{1/\alpha} \right)^{s-1}\int^\infty_0
d\xi \, \xi ^{s-1}\int^{+\infty }_{-\infty} d\eta  \, e^{i\eta \xi -|\eta |^\alpha }.  
\ee
The integrals over $d\xi $ and $d\eta  $ can be evaluated by using the
equation \cite{West2}:
%%%equations (3.3) and (3.6) of the Ref. \cite{West2}. Indeed, we have
\be \label{eq21}
\int^\infty_0 d\xi \, \xi^{s-1} \int^\infty_0 d\eta \, e^{i\eta  \xi -\eta^\alpha }=
\frac{4}{s-1} \sin \frac{\pi (s-1)}{2} \Gamma (s)\Gamma \Bigl(1-\frac{s-1}{\alpha} \Bigr),  
\ee
where $s-1<\alpha \leq 2$ and $\Gamma (s)$ is the Gamma function.

Inserting of (\ref{eq21}) into (\ref{eq20}) and using the relations 
\be
\Gamma (1-z)=-z\Gamma (-z) ,\quad \Gamma (z)\Gamma (1-z)=\pi /\sin \pi z, 
\ee
we find
\be \label{st}
\stackrel{\wedge }{K} (s,t )=\frac{1}{\alpha} 
\left( (C_\alpha t )^{1/\alpha} \right) ^{s-1} 
\frac{\Gamma (s)\Gamma ( \frac{1-s}\alpha )}{\Gamma (\frac{1-s}2)\Gamma (\frac{1+s}2)}. 
\ee
Then the inverse Mellin transform of (\ref{st}) is
\be
K(x,t )=\frac{1}{2\pi i}\int\limits_{c-i\infty }^{c+i\infty }dsx^{-s}%
\stackrel{\wedge }{K}(s,t )=
\frac{1}{2\pi i} \frac{1}{\alpha} \int\limits_{c-i\infty }^{c+i\infty}
ds \left( (C_\alpha t )^{1/\alpha } \right)^{s-1} 
x^{-s}\frac{\Gamma (s)\Gamma (\frac{1-s}\alpha )}{\Gamma (\frac{1-s} 2)\Gamma (\frac{1+s}2)} ,
\ee
where the integration contour is the straight line from $c-i\infty $ to 
$c+i\infty $ with $0<c<1$.  Replacing $s$ by $-s$, we get
\be \label{eq22}
K(x,t )= \frac{1}{\alpha} ( C_\alpha t )^{-1/\alpha} 
\frac{1}{2\pi i} \int\limits_{-c-i\infty }^{-c+i\infty}ds
\left( (C_{\alpha} t)^{-1/\alpha} x \right)^s 
\frac{\Gamma (-s)\Gamma (\frac{1+s}\alpha )}{\Gamma (\frac{1+s}2)\Gamma (\frac{1-s}2)}. 
\ee
The integration contour may be deformed into one running clockwise around 
$[-c,\infty)$. Comparison with the definition of the Fox $H$-function 
\cite{Fox,Mathai,Srivastava} gives
\be \label{eq23}
K(x,t )=  
\frac{1}{\alpha} \left( C_\alpha t \right)^{-1/\alpha} 
H_{2,2}^{1,1}\left[ (C_\alpha t )^{-1/\alpha} x \, \, \Bigl| \,  \,  
\frac{(1-1/\alpha ,1/\alpha ),(1/2,1/2)}{(0,1),(1/2,1/2)}\right] . 
\ee
Using the properties of the Fox $H$-function \cite{Fox,Mathai,Srivastava},
we obtain
\be \label{eq25}
K (x,t )=\frac{1}{\alpha |x|}H_{2,2}^{1,1}\left[ 
(C_\alpha t )^{-1/\alpha} |x| \, \, \Bigl| \,  \,  
\frac{(1,1/\alpha),(1,1/2)}{(1,1),(1,1/2)}\right] . 
\ee

%%%\section{Fox function representation of $K(x,t)$ for $\alpha=2$}

Let us show by analogy with \cite{Laskin} (see also \cite{Laskin2})
that Eq. (\ref{eq25}) includes as a particular case at $\alpha =2$
the well known Gauss distribution (\ref{Ka2}).
Assuming $\alpha =2$ in Eq. (\ref{eq25}), 
\be \label{KK2}
K(x,t )|_{\alpha =2}=
H_{2,2}^{1,1} \left[ (C_2t )^{-1/2} |x|\, \Bigl| \, \frac{(1,1/2),(1,1/2)}{(1,1),(1,1/2)} \right] .
\ee
The series expansion of the function (\ref{KK2}) gives
\be \label{KK}
K(x,t )|_{\alpha =2}=\frac{1}{2} \left( C_2 t \right)^{-1/2}
\sum\limits_{k=0}^\infty \left( - ( C_2t )^{-1/2}\right)^k 
\frac{|x|^k}{k!} \frac{1}{\Gamma (\frac{1-k}2)}. 
\ee
Substituting of $k\rightarrow 2l$ into (\ref{KK}), and using
\be \label{eq28}
\Gamma \Bigl(\frac{1}{2}-l\Bigr)=\frac{\sqrt{\pi}}{(-1)^l (2l)!} (2)^{2l}l! ,
\ee
the function $K(x,t)$ can be rewritten as
\be \label{eq29}
K(x,t )|_{\alpha =2} = \frac{( C_2t )^{-1/2}}{2\sqrt{\pi}} 
\sum\limits_{l=0}^\infty \left( - (C_2t )^{-1/2} \right)^{2l} 
\frac{(-1)^lx^{2l}}{ 2^{2l}l!}= 
\frac{1}{\sqrt{4\pi C_2t}} \exp \Bigl(- \frac{x^2}{4C_2t }\Bigr) . 
\ee
Thus, it is shown that (\ref{Ka2}) can be derived 
from equation (\ref{eq25}) with $\alpha=2$.

%%%%%%%%%%%%%%%%%%%%%%%%%%%%%%%%%%%%%%%%%%%%%%%%%%%%%%%%%%%%%%%

\end{document}